# Implication of Repatriating Migrant Workers on COVID-19 Spread and Transportation Requirements

Avijit Maji[a], Tushar Choudhari[b], M.B. Sushma[c]

[a] Department of Civil Engineering, Indian Institute of Technology Bombay, Mumbai - 400076, India, avijit.maji@gmail.com *(Corresponding Author)*

[b] Department of Civil Engineering, Indian Institute of Technology Bombay, Mumbai - 400076, India, tp.choudhari@gmail.com

[c] Department of Civil Engineering, Indian Institute of Technology Bombay, Mumbai - 400076, India, prustysushma618@gmail.com

## Highlights

- Lower daily arrival of migrant workers preferred for Uttar Pradesh and Bihar
- Need a balanced number of days to repatriate all migrant workers from all states
- Daily arrival rate of migrant workers does not affect Rajasthan after May 31, 2020
- Repatriating migrant workers have a minimum implication on Maharashtra and Rajasthan
- Highest number of trains are needed to repatriate migrant workers to Bihar

## Abstract

Nationwide lockdown for COVID-19 created an urgent demand for public transportation among migrant workers stranded at different parts of India to return to their native places. Arranging transportation could spike the number of COVID-19 infected cases. Hence, this paper investigates the potential surge in confirmed and active cases of COVID-19 infection and assesses the train and bus fleet size required for the repatriating migrant workers. The expected to repatriate migrant worker population was obtained by forecasting the 2011 census data and comparing it with the information reported in the news media. A modified susceptible-exposed-infected-removed (SEIR) model was proposed to estimate the surge in confirmed and active cases of COVID-19 patients in India's selected states with high outflux of migrants. The developed model considered combinations of different levels of the daily arrival rate of migrant workers, total migrant workers in need of transportation, and the origin of the trip dependent symptomatic cases on arrival. Reducing the daily arrival rate of migrant workers for states with very high outflux of migrants (i.e., Uttar Pradesh and Bihar) can help to lower the surge in confirmed and active cases. Nevertheless, it could create a disparity in the number of days needed to transport all repatriating migrant workers to the home states. Hence, travel arrangements for about 100,000 migrant workers per day to Uttar Pradesh and Bihar, about 50,000 per day to Rajasthan and Madhya Pradesh, 20,000 per day to Maharashtra and less than 20,000 per day to other states of India was recommended.

## 1. Introduction

The worldwide spread of novel coronavirus, COVID-19, is recognized as a pandemic by the World Health Organization (WHO). In India, till April 29, 2020, total 33,065 confirmed, 8,429 recoveries and 1,079 deaths were reported (MoHFW, 2020). Like other countries, the Government of India adopted social distancing as a non-pharmaceutical infection prevention and control intervention. As part of this strategy, the large-scale population movements were avoided or restricted since March 24, 2020. For this, a nationwide lockdown with a complete restriction on passenger travel by all transportation modes was implemented. In the absence of a vaccine, social distancing is argued as an effective prevention and control strategy (Singh and Adhikari, 2020; Chatterjee et al., 2020). However, lockdown, along with a complete restriction on passenger travel, has an immediate adverse effect on migrant workers, particularly for those who belong to the marginalized sections of the society and depend on daily wages for living. Approximately 10.55 million migrant workers were reported to be living in 22,567 shelters set up in various regions of India (Rawal et al., 2020). An incidence of mass gathering on April 14, 2020 (the last day of the first phase of lockdown) near railway stations at Mumbai, Maharashtra, and Surat, Gujarat signified the strong desire of the migrant workers to return to their native places (Joshi et al., 2020).

The spread of COVID-19 is associated with the infected patient's travel history (Kraemer et al., 2020; Chinazzi et al., 2020; Singhal, 2020). So, repatriating migrant workers could increase the risk of spreading COVID-19 not only among co-passengers but also within the communities in their native places. Since the first week of May 2020, the Government of India and various State Governments arranged special trains and buses for the stranded migrant workers (Banerjea, 2020; PTI, 2020b; PTI, 2020c). Multiple studies described the impact of lockdown on COVID-19 transmission in India (Singh and Adhikari, 2020; Chatterjee et al., 2020; Mukhopadhyay and Chakraborty, 2020). However, the consequence of repatriating migrant workers on the COVID-19 spread in their home state, and the transportation needs were not explored so far. Furthermore, there is no specific transportation model to assess the demand during pandemic from a contagious disease like COVID-19, particularly when meeting the demand have consequence in terms of COVID-19 infection spread. So, this paper attempts to estimate the possible implication in terms of the surge in the confirmed and active cases of COVID-19 infection in Indian states with a high influx of repatriating migrant workers. The estimation considered scenarios with different levels of the daily arrival rate of migrant workers, total migrant workers in need of transportation, and origin of trip dependent symptomatic migrant workers. A modified susceptible-exposed-infected-removed (SEIR) model was proposed to predict the temporal variation of confirmed and active cases in the selective Indian states. The study also assessed the train and bus fleet size for different daily arrival rates of migrant workers until May 31, 2020. The concerned agencies can use the results to schedule and arrange resources for efficient transportation of repatriating migrant workers.

## 2. SEIR Models

Mathematical modeling based on the differential dynamic formulations, such as susceptible-infected-removed (SIR) and SEIR models, provides comprehensive information on the transmission mechanism of the COVID-19 epidemic. There are different variants of deterministic SIR and SEIR models to study the outbreak of the epidemic diseases (Tang et al., 2020a; Quilty et al., 2019; Shen et al., 2020; Tang et al., 2020b; Nadim et al., 2019; Read et al., 2020). For example, a generalized SEIR model developed by Tang et al. (2020a and 2020b) and Chen et al. (2020) successfully evaluated the transmission risk and predicted the number of COVID-19 infected persons in China for different scenarios. Tang et al. (2020b) further extended the SEIR model to incorporate time-dependent contact and diagnosis rates.

At any point of time $t$, a base SEIR model, as shown in Figure 1, divides the population of a state or region ($N$) into seven different compartments, i.e., susceptible ($S$), exposed ($E$), infected ($I$), quarantined ($Q$), recovered ($R$), deceased ($D$) and insusceptible or isolated ($P$). The flow of population through these compartments was modeled using a set of differential equations given in Table 1. In this model, time-dependent functions were considered to represent the reduction in COVID-19 related death, increment in recovery, and faster identification of COVID-19 positive patients over time due to improvements and inventions in medical science. However, like any other SEIR model, it did not consider the births and natural deaths. So, this base SEIR model is useful to study the outbreak of the epidemic diseases in a specific region or state until the arrival of repatriating migrant workers from different regions or states. The base SEIR model was modified by adding time-dependent interacting compartments for the repatriating migrant workers. These compartments interacted with the base SEIR model compartments only for a specific time within the analysis period. The working principle of the base SEIR and modified SEIR models are described in the following sub-sections.

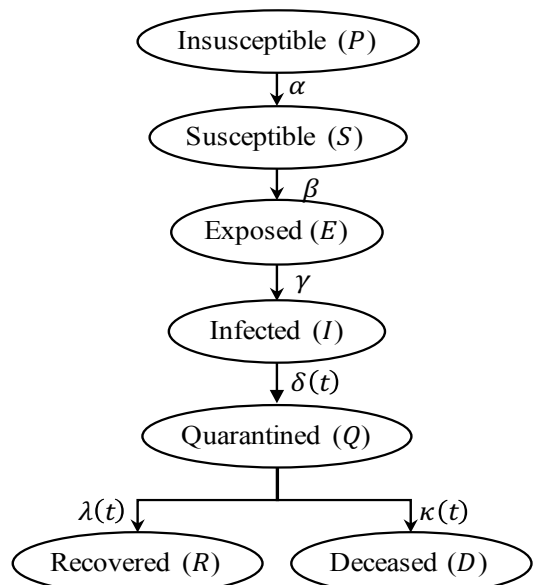

**Figure 1: Base SEIR model for COVID-19**

**Table 1: Differential equations for the compartmental base SEIR model**

| | |
|---|---|
| $\dfrac{dS(t)}{dt} = -\beta \dfrac{S(t)I(t)}{N} + \propto P(t)$ | $\dfrac{dE(t)}{dt} = \beta \dfrac{S(t)I(t)}{N} - \gamma E(t)$ |
| $\dfrac{dI(t)}{dt} = \gamma E(t) - \delta(t)I(t)$ | $\dfrac{dQ(t)}{dt} = \delta(t)I(t) - \lambda(t)Q(t) - \kappa(t)Q(t)$ |
| $\dfrac{dR(t)}{dt} = \lambda(t)Q(t)$ | $\dfrac{dD(t)}{dt} = \kappa(t)Q(t)$ |
| $\dfrac{dP(t)}{dt} = -\propto P(t)$ | $\delta(t) = \delta_1\left(1 - e^{-\delta_2 \times t}\right)$ |
| $\lambda(t) = \lambda_1\left(1 - e^{-\lambda_2 \times t}\right)$ | $\kappa(t) = \kappa_1 e^{-\kappa_2 \times t}$ |
| Where, $S + E + I + Q + R + D = N$ | |
| $N$ = Total population | $\delta(t)$ = Rate of detection leading to quarantine |
| $\propto$ = Rate of being susceptible | $\lambda(t)$ = Rate of recovery |
| $\beta$ = Rate of exposure | $\kappa(t)$ = Rate of mortality |
| $\gamma$ = Rate of being infected after exposure | $\delta_1, \delta_2, \lambda_1, \lambda_2, \kappa_1, \kappa_2$ = Constants |

## 2.1 Base SEIR model

At each time step of the base SEIR model (see Figure 1), the $P$ compartment population, when defying isolation for various reasons, became susceptible at the rate of $\alpha$ and moved to the $S$ compartment. Those in the $S$ compartment, got exposed (when in contact with an infected individual) at the rate of $\beta$ and eventually got infected with COVID-19 at the rate of $\gamma$. Accordingly, they were removed from the $S$ compartment and added to the $E$ compartment, and subsequently to the $I$ compartment. The removal rate of $\gamma$ from the $E$ compartment was equivalent to 1/incubation period or the latent time. Depending on the symptom and test protocol, certain people of the overall population were tested. If the tested person belonged to the $I$ compartment, the test result was expected to be positive. These individuals were either hospitalized or isolated in an institutional quarantine center, which was represented by the $Q$ compartment. Individuals from the $Q$ compartment either got cured or recovered and moved to the $R$ compartment at $\lambda(t)$ rate or died and moved to the $D$ compartment at the mortality rate of $\kappa(t)$. Since all the infected individuals could not be identified without testing the entire population, some, particularly the asymptomatic individuals, would keep spreading the virus. The population in each compartment of a base SEIR model at time $t$, i.e., $S(t)$, $E(t)$, $I(t)$, $Q(t)$, $R(t)$, $D(t)$ and $P(t)$, and the parameters, i.e., $\alpha, \beta, \gamma, \delta(t), \lambda(t)$ and $\kappa(t)$, were derived by solving the ordinary differential equations shown in Equation 1 for the historical data of confirmed, active, quarantined, etc. cases. So, the proficiency of an SEIR model greatly depends on the historical data. Open-access MatLab code developed by Cheynet (2020) was suitably modified to solve Equation 1 using the standard fourth-order Runge-Kutta method. In this model, the total population was considered as $N_p + MW_p$, where $N_p$ represents the total population of a region or state and $MW_p$ the expected repatriating migrant population. The $N_p$ was distributed among the $S, E, I, Q, R, D$ and $P$ compartments, and the $MW_p$ remained as a non-interacting population in the base SEIR model.

$$\frac{d[S\ E\ I\ Q\ R\ D\ P]^T}{dt} = \begin{bmatrix} 0 & 0 & 0 & 0 & 0 & 0 & \alpha \\ 0 & -\gamma & 0 & 0 & 0 & 0 & 0 \\ 0 & \gamma & -\delta & 0 & 0 & 0 & 0 \\ 0 & 0 & \delta & -\lambda-\kappa & 0 & 0 & 0 \\ 0 & 0 & 0 & \lambda & 0 & 0 & 0 \\ 0 & 0 & 0 & \kappa & 0 & 0 & 0 \\ 0 & 0 & 0 & 0 & 0 & 0 & -\alpha \end{bmatrix} \begin{bmatrix} S(t) \\ E(t) \\ I(t) \\ Q(t) \\ R(t) \\ D(t) \\ P(t) \end{bmatrix} + \begin{bmatrix} -\beta \\ \beta \\ 0 \\ 0 \\ 0 \\ 0 \\ 0 \end{bmatrix} \frac{S(t)I(t)}{(N_p + MW_p)} \quad (1)$$

## 2.2 Modified SEIR model for repatriating migrant worker

On arrival, migrant workers were not tested for COVID-19 infection, but only symptoms were monitored. Similarly, at the time of boarding a train or bus, only symptoms were monitored. Hence, asymptomatic and exposed individuals could still board a train or bus, infect co-passengers, and become symptomatic before alighting. The probability of a migrant worker being symptomatic on arrival at home state depended on their last residence or state before the trip. If the probability of being symptomatic and quarantined, estimated by considering the ratio of quarantined population and the total population, of a particular state $i$ is $QR_i$, the expected number of migrant workers detected symptomatic on arrival among the migrant worker population of $MW_{ij}$ traveling from state $i$ to state $j$ was $QR_i \times MW_{ij}$. In a confined compartment of a train or bus, an infected individual can expose other co-passengers. Hence, a high value of $QR_i$ was considered to estimate the infected number of migrant workers reaching state $j$ from state $i$. Based on the Indian Council of Medical Research (ICMR) report, the number of asymptomatic cases varied between 50 to 82%, with an average of 69% (Chauhan, 2020). Other studies also reported the ratio of the asymptomatic and symptomatic population as two-third or higher (Nishiura et al., 2020; Mizumoto et al., 2020). So, the expected number of asymptomatic migrant workers arriving at home state was assumed to be about two-thirds of the number of symptomatic migrant workers. The symptomatic migrant workers would be quarantined (in hospital or quarantine center) at their home state. The co-passengers from the same compartment of symptomatic passengers might be quarantined as well. All other migrant workers were expected to be in isolation for 14 days. The assumed situations of migrant workers at their origin and destination states are shown in Figure 2.

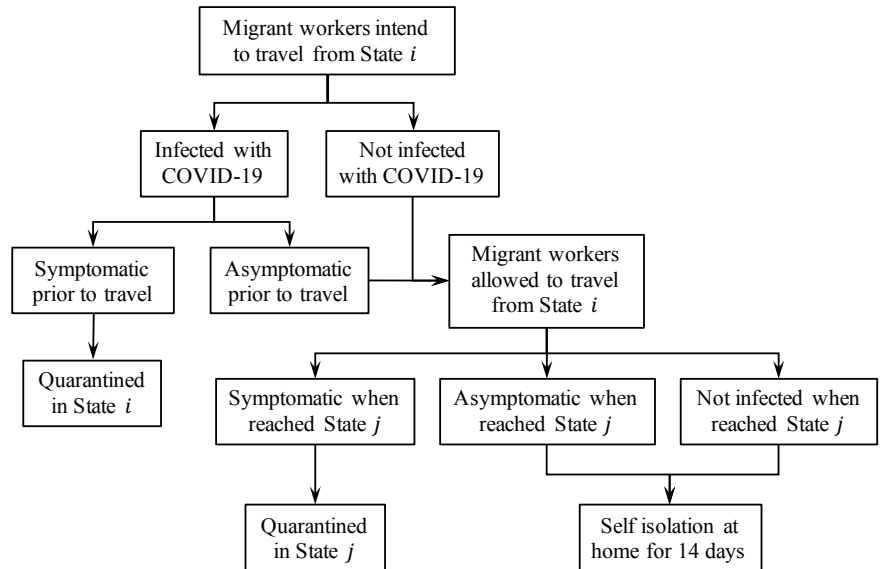

**Figure 2: Situation of migrant workers at origin and destination**

An interaction was established between a base SEIR and the $MW_p$ when the migrant workers started arriving. It helped to assign the repatriating migrant workers to the appropriate compartments. For example, the symptomatic migrant workers were assigned to the $Q$ compartment, asymptomatic to the $I$ compartment, and remaining to the $P$ compartment. The modified SEIR model with possible interactions for the arriving migrant workers is shown in Figure 3. Note that the migrant worker population did not interact with any compartments of the base SEIR model until the first arrival day, and hence, no such interaction is shown in Figure 1. Total time required to transport all migrant workers to their respective home states would depend on the travel arrangements. Hence, this study adopted different levels of daily arrival rates of repatriating migrant workers for analyses.

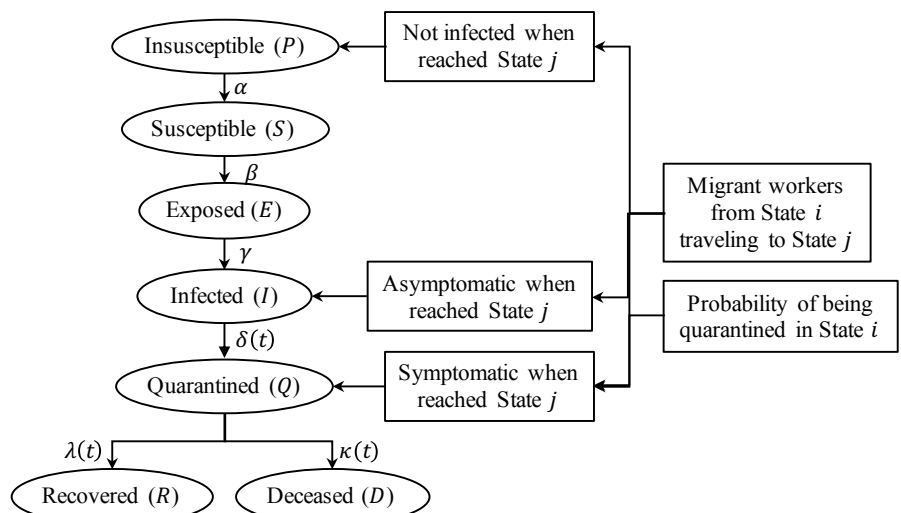

**Figure 3: Modified SEIR model**

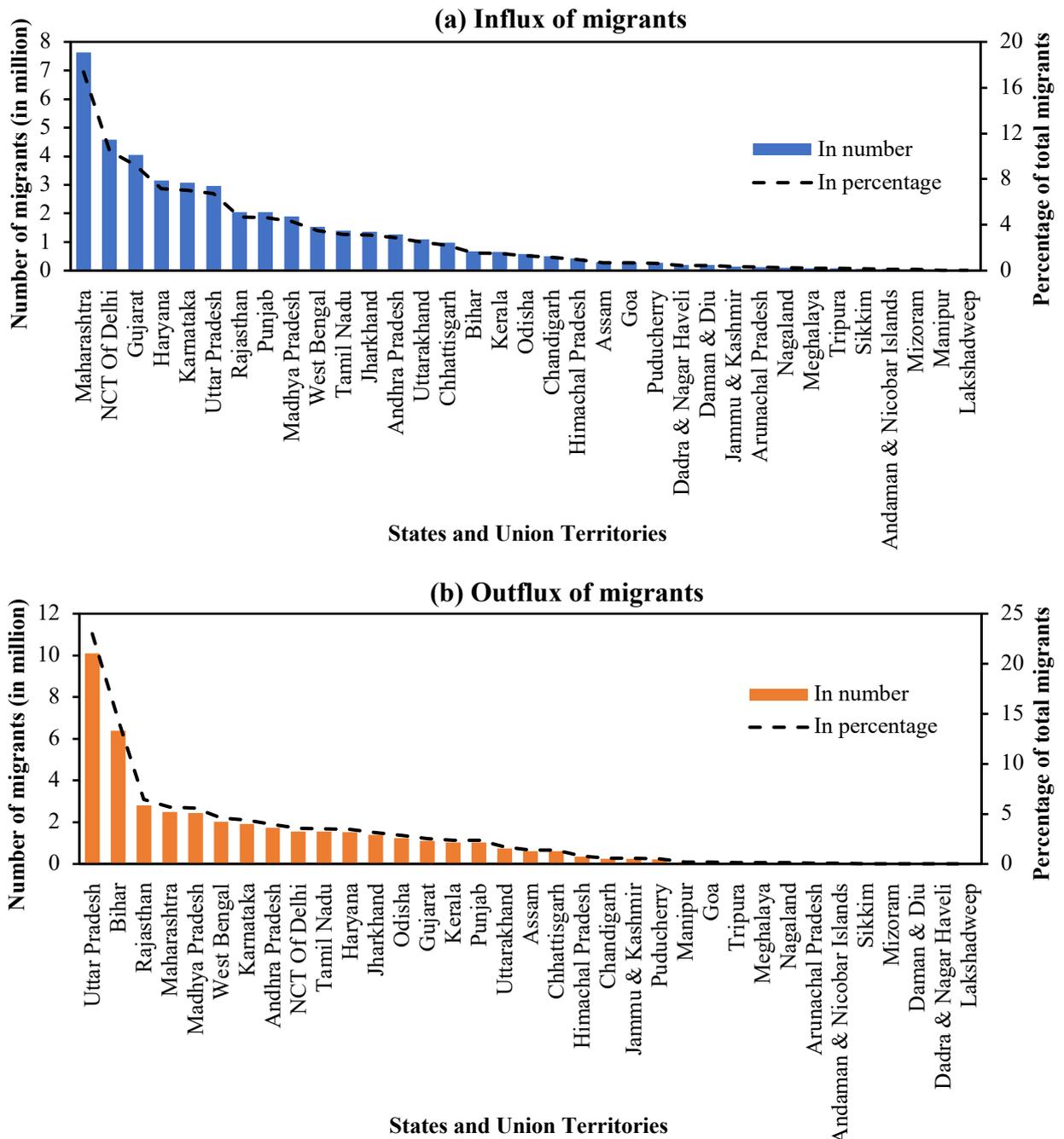

**Figure 4: Influx and outflux of migrants from 2011 census data**

## 3. Data Preparation and Analyses

In India, census data is collected every ten years. The last census data was collected in 2011, and the next census data would be collected in 2021. Hence, the 2011 census data published by the Office of the Registrar General & Census Commissioner, India (MoHA, 2020) was analyzed to identify states with the significantly high influx and outflux of the inter-state migrant population (includes migration for work, business, marriage, education, and other purposes). As observed in Figure 4, Maharashtra, National Capital Territory (NCT) of Delhi, Gujarat, Haryana and Karnataka had a higher influx of migrants (i.e., the states where people migrated to), and Uttar Pradesh, Bihar, Rajasthan, Maharashtra and Madhya Pradesh had a higher outflux of migrants (i.e., the states from where people migrated). Cumulatively these states contributed more than 50% of the influx and outflux of inter-state migrant population of India. So, in this study, Uttar Pradesh, Bihar, Rajasthan, Maharashtra, and Madhya Pradesh were analyzed for the repatriating migrant workers from Maharashtra, NCT of Delhi, Gujarat, Haryana, Karnataka, and other states and union territories (UT) of India. It is to be noted that Maharashtra had the highest influx of migrants and the fourth-highest outflux of migrants. As of May 15, 2020, Maharashtra, Gujarat, NCT of Delhi, Rajasthan, Madhya Pradesh, and Uttar Pradesh were among the top ten COVID-19 affected states and UTs of India.

Further, the 1981 to 2011 census data were used to obtain the $N_p$ and $MW_p$ for the analysis year 2020. There are different methods such as arithmetic increase, geometric increase, incremental increase, decrease in growth rate and logistic curve, to forecast population based on the land-use and economic status of a region (Jensen, 1995; Moalafhi et al., 2012; Lal, 2020; Jain et al., 2015). While predicting the 2011 population from the previous census data, the accuracy of the incremental increase method was found to be better than the other methods. Hence, Equation 2, representing the forecasted population by incremental increase method, was used to obtain the analysis year population. See Table 2 for the obtained population of the states considered.

$$N_p^n = N_p^l + \frac{n}{10}\bar{p} + \frac{n(n+10)}{200}\bar{q} \tag{2}$$

where,

$N_p^n$ = Population at $n^{th}$ year from last census data
$N_p^l$ = Population in last census data
$n$ = Number of years beyond the last census data
$\bar{p}$ = Average rate of increase in decadal population
$\bar{q}$ = Average rate of change in increase in decadal population

**Table 2: Predicted population**

| States | Population (in million) | | | | | $\bar{p}$ (in million) | $\bar{q}$ (in million) |
|---|---|---|---|---|---|---|---|
| | 1981 | 1991 | 2001 | 2011 | 2020 | | |
| India | 685.18 | 838.58 | 1028.74 | 1210.19 | 1568.02 | 175.00 | 14.03 |
| Maharashtra | 62.78 | 78.94 | 96.75 | 112.37 | 127.02 | 16.53 | -0.27 |
| NCT of Delhi | 6.22 | 9.42 | 13.85 | 16.75 | 19.79 | 3.51 | -0.15 |
| Gujarat | 34.09 | 41.31 | 50.67 | 60.44 | 69.43 | 8.78 | 1.27 |
| Haryana | 12.92 | 16.46 | 21.14 | 25.35 | 29.36 | 4.14 | 0.33 |
| Karnataka | 37.14 | 44.98 | 52.85 | 61.10 | 68.46 | 7.99 | 0.20 |
| Uttar Pradesh | 110.86 | 139.11 | 166.20 | 199.81 | 228.79 | 29.65 | 2.68 |
| Bihar | 69.91 | 64.53 | 82.10 | 104.10 | 125.68 | 11.39 | 13.24 |
| Rajasthan | 34.26 | 44.01 | 56.51 | 68.55 | 79.82 | 11.43 | 1.15 |
| Madhya Pradesh | 52.18 | 48.57 | 60.35 | 72.63 | 85.55 | 6.82 | 7.95 |

The input parameters for the modified SEIR model were $\alpha, \beta, \gamma, \delta(t), \lambda(t), \kappa(t)$, daily arrival rate of migrant workers, total number of repatriating migrant workers and their infection probability. The $\alpha, \beta, \gamma, \delta(t), \lambda(t)$ and $\kappa(t)$ for the modified SEIR model were obtained from the base SEIR model. However, at the time of the analysis, reliable statistics on the total number of repatriating migrant workers, their daily arrival rate, and infection probability were not available. In such a situation, exploratory modeling and analysis (EMA) could use computational experiments to analyze uncertain and complex conditions (Kwakkel and Pruyt, 2012). This approach develops carefully selected scenarios by appropriate assumptions/guesses to exploit the relevant information. A similar process was adopted in this study to utilize the available information related to the migrant population and infection probability and simulate various scenarios using the modified SEIR model.

*Implication of Repatriating Migrant Workers on COVID-19 Spread and Transportation Requirements*

### 3.1 Expected repatriating migrant workers

Migrant workers residing for less than four years could be socially and economically vulnerable. Therefore, the number of migrant workers expected to repatriate was derived from Equation 3 based on the number of migrants arriving in the last four years of the census year. The forecasted total influx and outflux of migrants in the previous four years of analysis year are presented in Table 3, and the overall process of forecasting is shown in Figure 5. About 8-14% of the forecasted outflux migrants of Uttar Pradesh, Bihar, Madhya Pradesh, and Rajasthan, and about 4-7% of the forecasted outflux migrants of Maharashtra was assumed to repatriate. Among all the Indian states and UTs, Maharashtra had the highest number of confirmed patients and death due to COVID-19. The migrant workers may refrain from returning to Maharashtra (home state). Hence, a lower repatriation range was adopted for Maharashtra. As shown in Table 4, the number of migrant workers in scenarios 1, 2, and 3 were arbitrarily chosen from the lower, middle, and upper terciles (3-quantiles) of the range of repatriating migrant workers, respectively. Information from various published sources, as presented in Table 3, confirmed the repatriating migrant workers as 1.9-21% (with an average of about 10.2%) of the influx or outflux migrants recorded in the last four years.

$$M_p^n = N_p^n \times r_{avg}^{4/9} \times (1 + \overline{gr})^n \qquad (3)$$

where,

$M_p^n$ = Migrant population in the previous four years at $n^{th}$ year
$\overline{gr}$ = Average decadal growth rate of $M_p/N_p$ ratio
$r_{avg}^{4/9}$ = Average ratio of migrant population in the previous four and nine years

Table 3: Predicted migrants and migrant workers

| States | Migrants in last four years (in millions) | | Number of repatriating migrant workers reported in news media (in thousands) | | | Percentage of predicted influx or outflux migrants |
|---|---|---|---|---|---|---|
| | Influx | Outflux | Leaving | Arriving | Reference | |
| Maharashtra | 9.12 | 2.79 | 650 | *NA* | Bose (2020) | 7.1 |
| NCT of Delhi | 3.80 | - | 800 | *NA* | Mathur and Yadav (2020) | 21.0 |
| Gujarat | 5.79 | - | *NA* | *NA* | - | - |
| Haryana | 3.64 | - | 70 | *NA* | ET Bureau (2020) | 1.9 |
| Karnataka | 4.10 | - | 360 | *NA* | Kumar (2020a) | 8.8 |
| Uttar Pradesh | - | 10.87 | *NA* | 1,000 | PTI (2020a) | 9.2 |
| Bihar | - | 7.49 | *NA* | 1,000 | Kumar (2020b) | 13.4 |
| Rajasthan | - | 3.09 | *NA* | *NA* | - | - |
| Madhya Pradesh | - | 2.88 | *NA* | *NA* | - | - |

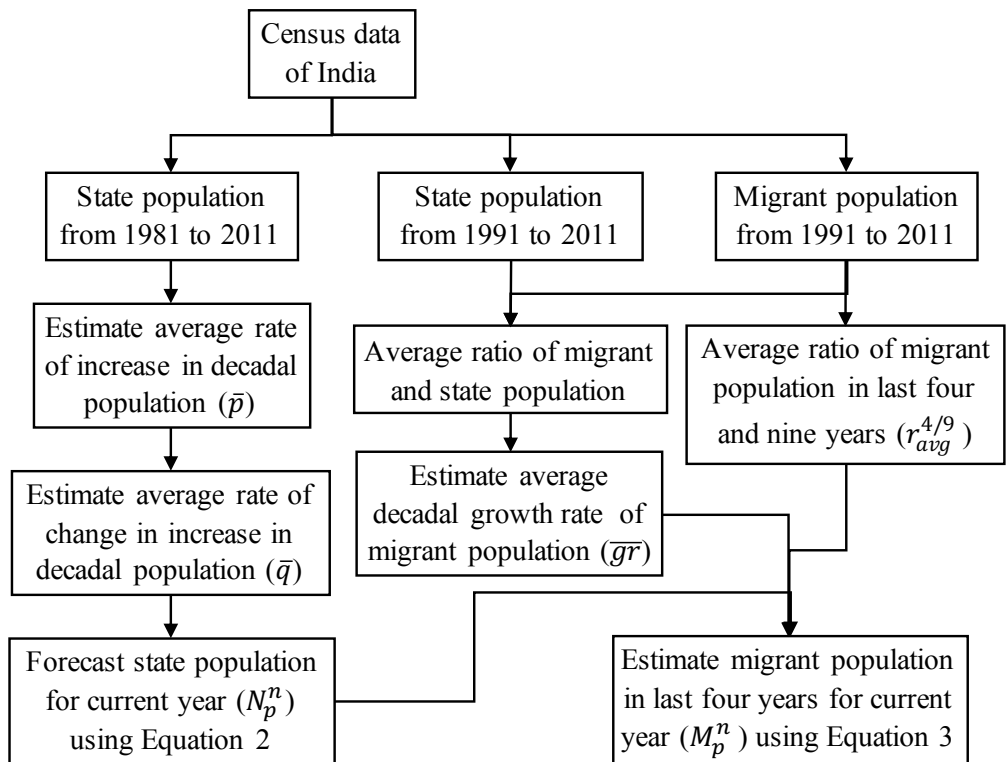

**Figure 5: Forecasting of state and migrant population**

Table 4: Scenarios related to total repatriating migrant workers

| Scenarios | Total repatriating migrant workers (in thousands) | | | | |
|---|---|---|---|---|---|
| | **Maharashtra** | **Uttar Pradesh** | **Bihar** | **Rajasthan** | **Madhya Pradesh** |
| Scenario 1 | 126 | 936 | 676 | 280 | 269 |
| Scenario 2 | 157 | 1169 | 845 | 350 | 337 |
| Scenario 3 | 189 | 1403 | 1014 | 420 | 404 |

### 3.2 Arrival rate of repatriating migrant workers

The expected repatriating migrant workers would arrive in phases due to limited public transportation facilities. Trains for ferrying migrant workers had a capacity of 1200 to 1700 passengers (Dhingra, 2020; Saxena, 2020), and buses were expected to transport at 50% capacity (Express News Service, 2020; Malik, 2020). Total number of trains and buses available for this purpose was about 300 and 10,000, respectively (Dhingra, 2020; Malik, 2020), which varied over time for different reasons (Dhingra, 2020; PTI, 2020b; PTI, 2020c; Saxena, 2020; Banerjea, 2020). Hence, this study adopted different daily arrival rates of repatriating migrant workers for analyses. The trains were highly preferred for interstate travel as buses traveled till state borders. The trains were expected to increase in number for fast mobilization (Dhingra, 2020). If 5-10% (15-30 trains) of the 300 trains operating throughout India are traveling daily to Uttar Pradesh (highest outflux state), about 18,000 to 36,000 passengers can reach Uttar Pradesh per day. Therefore, this study explored the daily arrival rates between 20,000 and 100,000 migrant workers (at an interval of 10,000 migrant workers per day) for the three scenarios shown in Table 4.

### 3.3 Infection probability of repatriating migrant workers

Social distancing requirements are compromised in the public transportation mode. Hence, COVID-19 infection increases significantly with the use of public transport (Zheng et al., 2020). During repatriation, migrant workers would travel 12 to 72 hours in a confined space of a train or bus, and use the common facilities such as washroom, train station, bus station, and *Implication of Repatriating Migrant Workers on COVID-19 Spread and Transportation Requirements*catering service. Therefore, this study explored the infection probability between 10 and 100 times (at an interval of 30 times) of the average infection probability of the trip originating state for the three scenarios shown in Table 4.

The base SEIR model was simulated once for a state to obtain the base scenario with no effect of repatriating migrant workers. Whereas, the modified SEIR model was simulated 108 times to represent the effect of three scenarios of the total expected repatriating migrant workers for nine different daily repatriation rates and four possible infection probability (i.e., 3×9×4 = 108). The daily "confirmed", "recovered" and "death" related data available from March 14 to April 29, 2020, at https://www.covid19india.org (COVID-19 Tracker, 2020) was used in the base SEIR model. A minimum of ten confirmed cases was considered to initiate the analysis. Hence, for specific states like Madhya Pradesh and Bihar with the late spread of infection, the analysis start date was after March 14, 2020. The probability of being symptomatic and quarantined estimated based on the active and quarantined population of a state on April 29, 2020 and the total population, and the four possible infection probabilities (i.e., Options A, B, C, and D) are presented in Table 5. The distribution of repatriating migrant workers between origin and home states, estimated from the migration details available in 2011 census data, is shown in Table 6.

Table 5: Probability of being symptomatic and quarantined

| States | Active and quarantined cases as of April 29, 2020 | | Some of the options with elevated values of $QR_i$ (per million population) | | | |
|---|---|---|---|---|---|---|
| | Total | Per million population ($QR_i$) | Option A: 10 times | Option B: 40 times | Option C: 70 times | Option D: 100 times |
| India | 23546 | 15.02 | 150.16 | 600.66 | 1051.15 | 1501.64 |
| Maharashtra | 7890 | 62.12 | 621.16 | 2484.65 | 4348.13 | 6211.62 |
| NCT of Delhi | 2291 | 115.77 | 1157.66 | 4630.62 | 8103.59 | 11576.55 |
| Gujarat | 3358 | 48.37 | 483.65 | 1934.61 | 3385.57 | 4836.53 |
| Haryana | 83 | 2.83 | 28.27 | 113.08 | 197.89 | 282.70 |
| Karnataka | 297 | 4.34 | 43.38 | 173.53 | 303.68 | 433.83 |
| Uttar Pradesh | 1585 | 6.93 | 69.28 | 277.112 | 484.95 | 692.78 |
| Bihar | 337 | 2.68 | 26.81 | 107.26 | 187.70 | 268.14 |
| Rajasthan | 1569 | 19.66 | 196.56 | 786.27 | 1375.97 | 1965.67 |
| Madhya Pradesh | 1969 | 23.02 | 230.16 | 920.63 | 1611.11 | 2301.58 |
| Rest of India | 4167 | 5.68 | 56.76 | 227.05 | 397.33 | 567.62 |

Table 6: Distribution of repatriating migrant workers

| Migrated states | Repatriating migrant worker to home states (%) | | | | |
|---|---|---|---|---|---|
| | Maharashtra | Uttar Pradesh | Bihar | Rajasthan | Madhya Pradesh |
| Maharashtra | 0 | 26 | 11 | 18 | 30 |
| NCT of Delhi | 2 | 14 | 11 | 5 | 4 |
| Gujarat | 38 | 14 | 10 | 29 | 16 |
| Haryana | 2 | 10 | 8 | 14 | 4 |
| Karnataka | 21 | 2 | 3 | 6 | 2 |
| Uttar Pradesh | 3 | 0 | 11 | 6 | 14 |
| Bihar | 1 | 1 | 0 | 0 | 0 |
| Rajasthan | 4 | 5 | 2 | 0 | 15 |
| Madhya Pradesh | 10 | 5 | 1 | 9 | 0 |
| Rest of India | 20 | 22 | 43 | 13 | 16 |
| *Total* | *100* | *100* | *100* | *100* | *100* |

## 4. Numerical simulation results and discussion

The repatriating migrant workers were assumed to arrive in the respective states from May 5, 2020. Hence, in the modified SEIR model, the time-dependent daily interaction of the repatriating migrant worker compartment with the *I*, *Q* and *P* compartments (refer Section 2.2) were initiated from May 5, 2020. It remained active until the aggregated daily arrival reached the total expected repatriating migrant workers. Both the base and modified SEIR models were initiated by assigning the start day data to the corresponding recovered, deceased and quarantined compartments, and simulated until May 31, 2020. About 2000 times of the total quarantined, recovered, and deceased cases were assigned to the *S* compartment, the remaining state population without the repatriating migrant workers to the *P* compartment, and the repatriating migrant workers in the migrant worker compartment.

### 4.1 Parameter values

The $\alpha, \beta, \gamma, \delta_1, \delta_2, \lambda_1, \lambda_2, \kappa_1$ and $\kappa_2$ values for the base scenario are shown in Table 7, and its fitted trends for confirmed, active, recovered and deceased cases are shown in Figure 6. For all the states except Bihar, the $\alpha$ value, representing the rate of being susceptible, indicates that about 10% of the population can become susceptible. In Bihar, this rate is less (about 2%). Similarly, for all the states except Uttar Pradesh, the $\beta$ values indicated that the susceptible people were immediately exposed to the infection. It supports the high contagious characteristics of COVID-19 (Peng et al., 2020; Mangoni and Pistilli 2020). However, for Uttar Pradesh, this rate is low and may need further investigation for a better understanding of the interactions between people. The $\gamma$ value, indicating the rate of being infected after exposure, is almost similar for all the states analyzed. In other words, the incubation time ($1/\gamma$) or latent time of COVID-19 infection is about two days. The other parameters are state-specific, and depend on the confirmed, active, recovered and deceased cases recorded. The RMSE values presented in the corresponding plot (see Figure 6) indicate that the fitted model had comparatively higher errors for Madhya Pradesh and Rajasthan.

**Table 7: Model parameters from base SEIR model**

| Parameters | Values | | | | |
|---|---|---|---|---|---|
| | **Maharashtra** | **Uttar Pradesh** | **Bihar** | **Rajasthan** | **Madhya Pradesh** |
| $\alpha$ | 0.1 | 0.1 | 0.021 | 0.1 | 0.1 |
| $\beta$ | 0.92 | 0.5 | 1.2 | 0.9 | 1.0 |
| $\gamma$ | 0.6 | 0.6 | 0.4 | 0.6 | 0.6 |
| $\delta_1$ | 1.0 | 1.0 | 0.18 | 1.0 | 1.0 |
| $\delta_2$ | 0.039 | 0.015 | 0.1 | 0.045 | 0.055 |
| $\lambda_1$ | 0.021 | 1.0 | 0.043 | 1.0 | 1.0 |
| $\lambda_2$ | 0.066 | 0 | 0.1 | 0.0012 | 0.0009 |
| $\kappa_1$ | 0.028 | 0.002 | 0.0006 | 0.0016 | 0.0054 |
| $\kappa_2$ | 0.046 | 0 | 0 | 0 | 0 |

### 4.2 Surge in confirmed and active cases

The expected confirmed and active cases for the 108 scenarios were adjusted for the base scenario to estimate the surge in confirmed and active cases at each state. The range between the first and third quartiles and the median values (i.e., the second quartile) of these increased confirmed and active cases are presented in Figure 7. With stringent measures, the surge in confirmed and active cases could be within the first quartile. However, in the worst situation, the surge in confirmed and active cases could be higher than the third quartile. For a daily arrival rate of 20,000 migrant workers, Figures 7a and 7b represent the surge in confirmed and active cases, respectively. As expected, the confirmed cases in all the states were increasing during the analysis period (see Figure 7a). However, as shown in Figure 7b, the surge in active cases for some states such as Rajasthan and Uttar Pradesh were either flattening or reducing in the last week of May 2020. In these two states, the effect of repatriating migrant workers would be maximum around May 15, 2020 for the best-case scenario and around May 27, 2020, for the worst-case scenario.

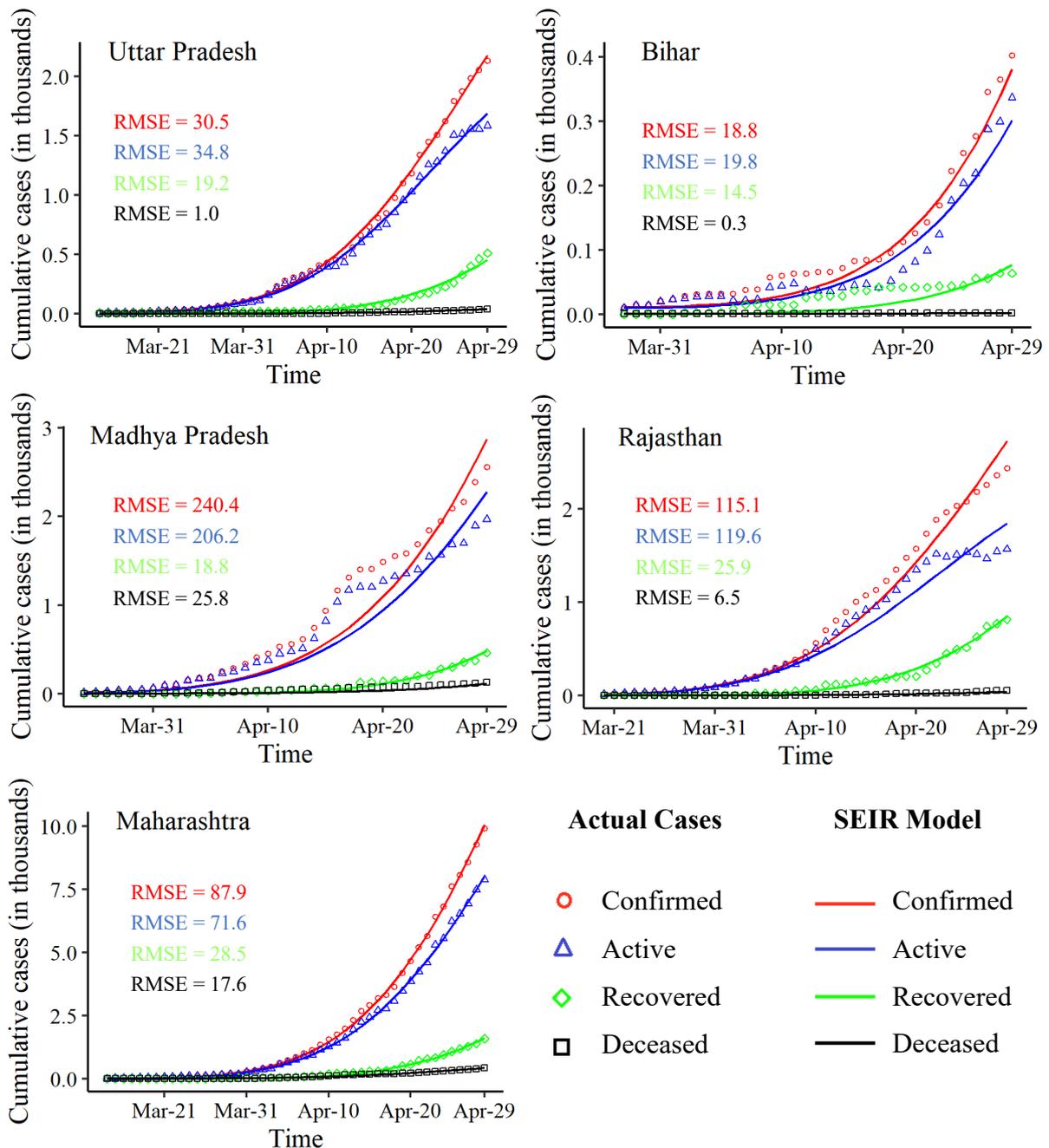

**Figure 6: Actual data and fitted trends of base SEIR model till April 29, 2020**

If the daily arrival rate of migrant workers increases to 100,000, among the states analyzed, Uttar Pradesh will experience the most significant surge in the confirmed and active cases at the beginning (see Figures 7c and 7d), which is expected to drop later (see Figure 7d). In the best-case scenario, Uttar Pradesh would experience a peak around May 16, 2020, and in the worst-case scenario, around May 23, 2020. A similar trend was observed for Rajasthan as well. In the best-case situation, Rajasthan would experience a peak around May 15, 2020, and in the worst-case scenario, around May 20, 2020. Irrespective of the daily arrival rate of migrant workers, Madhya Pradesh might experience a surge in confirmed and active cases until the end of the analysis period. The effect of migrant workers returning to Maharashtra was low compared to all the states analyzed. If the migrant workers leaving a state were adjusted, the surge in confirmed and active cases in Maharashtra was expected to be even lower than the cases presented in this paper.

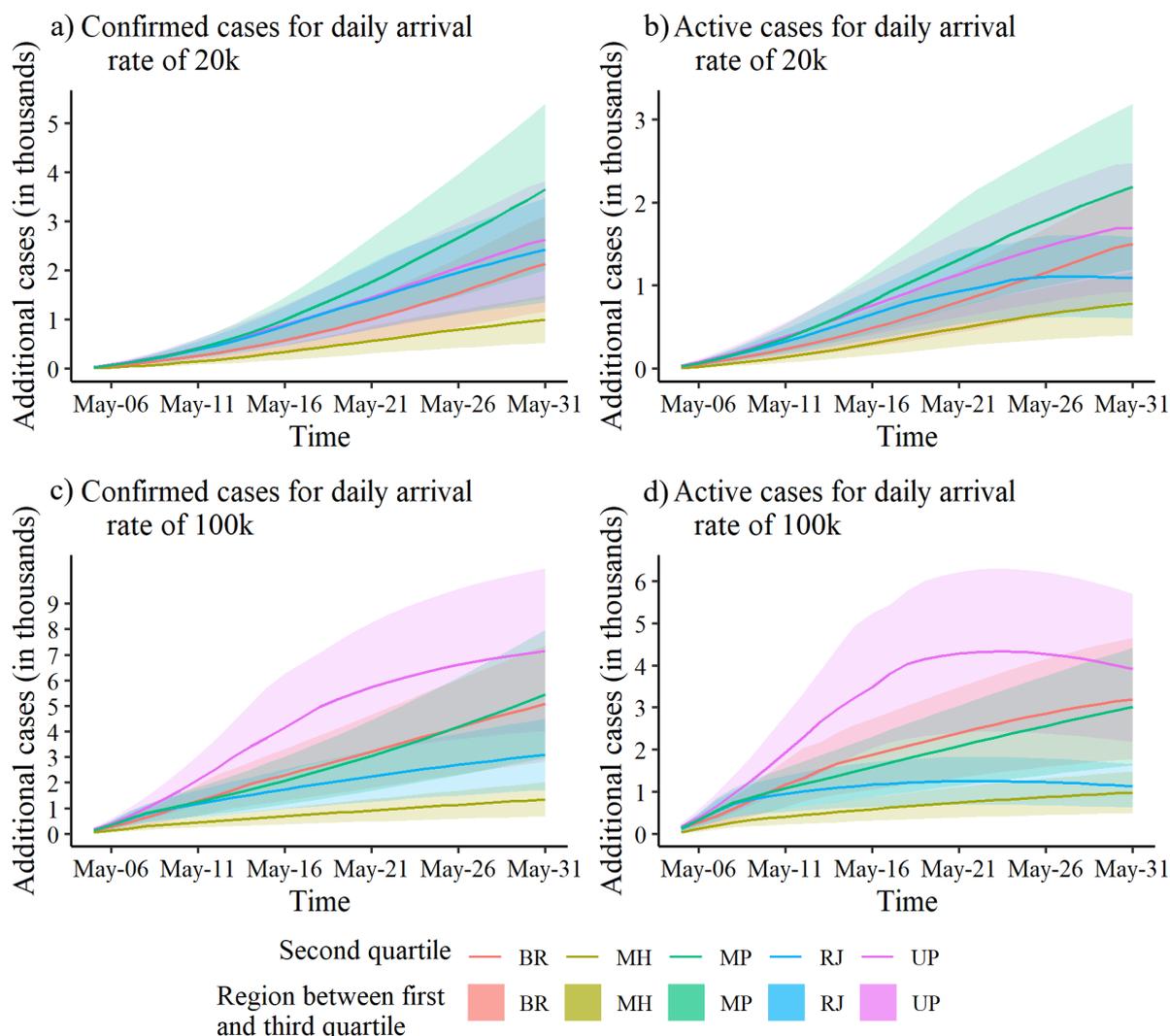

**Figure 7: Expected surge in confirmed and active cases till May 31, 2020**

The mean values of the surge in active cases for daily arrival rates of 20,000, 40,000, 60,000, 80,000, and 100,000 migrant workers are presented in Figure 8. It can be observed that for a daily arrival rate of 20,000 migrant workers, the surge in active cases in Uttar Pradesh and Bihar was considerably less. Further, by mid-May 2020, the surge in active cases for a daily arrival rate of 60,000 or more migrant workers was two to three times higher than the surge in active cases for the daily arrival rate of 20,000 migrant workers. Irrespective of the daily arrival rates, the surge in active cases in Rajasthan could be almost the same by the end of the analysis period. In other words, for Rajasthan, the effect of daily arrival rate on the active cases diminishes beyond May 31, 2020. However, such a trend was not observed for Uttar Pradesh, Bihar, Madhya Pradesh, and Maharashtra.

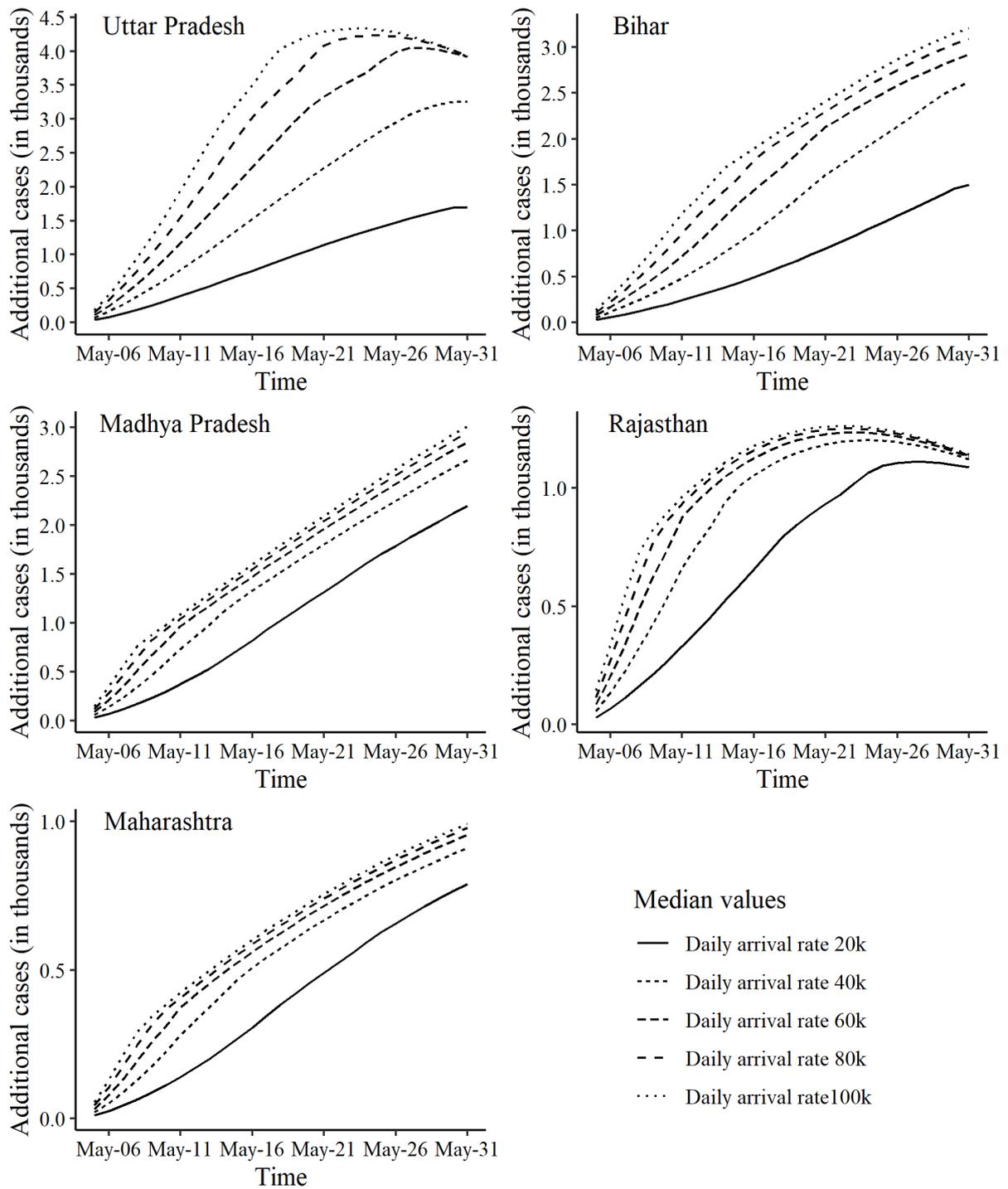

**Figure 8: Surge in active cases for different daily arrival rates**

## 5. Train and bus fleet size assessment

The travel distance, travel time, travel mode speed, travel mode capacity, expected total repatriating migrant workers indicated in Table 4, and distribution of repatriating migrant workers given in Table 6 were considered to establish the train and bus fleet size required to transport the repatriating migrant workers. It was assumed that 10% of the repatriating migrant workers would choose unorganized transportation modes such as hired private vehicles and non-motorized transport modes, and 90% would choose the organized transportation mode arranged by the State Governments and the Central Government. Buses arranged by the State Governments would ply up to the political boundaries of respective states. If a home and migrated states (or UTs) share a common border, the buses of migrated state would transport the repatriating migrant workers until the border of the home state. From there, the buses of the home state would transport them to their final destination. In this, the roundtrip travel time of a bus would approximately reduce to half; but, the number of buses required would double. It requires bilateral agreement or understanding between the migrated and home states. So, transportation of repatriating migrant workers may not be possible between states, not sharing a common border. In the presence of bus services, it was assumed that 45% of the repatriating migrant workers would travel by bus and the remaining 45% by train. Otherwise, 90% of the repatriating migrant workers would travel by train. The average speed of buses and trains were considered as 50 kmph and 80 kmph, respectively. Also, the roundtrip journey, disinfection time, cleaning time, average delay, time to change train engine, and refueling and driver break time for buses were appropriately assumed. The bus and train capacities were adopted as 25 (at 50% occupancy) and 1200, respectively. The fleet size assessed based on these assumptions is shown in Table 8, and the details regarding the estimation process in the Appendix.

Though the total number of migrant workers returning to Bihar was lower than Uttar Pradesh (see Table 4), the number of trains required to transport them to Bihar was marginally higher than the Uttar Pradesh. It is because Bihar is comparatively far from the highly migrated states (such as Maharashtra, NCT of Delhi, Gujarat, and Haryana) than Uttar Pradesh (see Table A1) and none of the highly migrated states have a common border with Bihar. Though Rajasthan and NCT of Delhi are well connected by road via Haryana, no buses were considered to transport migrant workers from NCT Delhi to Rajasthan. About 69% of the migrant workers repatriating to Maharashtra were expected from the three neighboring states of Gujarat, Karnataka, and Madhya Pradesh. Hence, for Maharashtra, the estimated bus fleet size is relatively large, even though the total repatriating migrant workers were comparatively low (see Table 4).

Table 8 revealed that at a daily arrival rate of 20,000, more than a month is needed to ferry the migrant workers to Uttar Pradesh and Bihar. It is a considerably longer duration than the time required to repatriate migrant workers to states like Rajasthan, Madhya Pradesh, and Maharashtra. Arranging transportation for about 100,000 migrant workers per day to Uttar Pradesh and Bihar, about 50,000 to Rajasthan and Madhya Pradesh, 20,000 to Maharashtra and less than 20,000 to other states of India can bring some parity in the total number of days required to transport the repatriating migrant workers. The decision on the daily arrival rate of migrant workers also depends on the transportation and health infrastructure available in the receiving states. For example, Bihar was unable to facilitate more buses for repatriating migrant workers (FE Online, 2020). Similarly, Figure 8 indicates the increased need for health infrastructure in Uttar Pradesh and Bihar for the higher daily arrival rates. The transportation operating agencies also need to consider infrastructure capacity and availability of the human resources in deciding the train and bus fleet size. For example, the Indian Railways could operate only 115 "Shramik Trains" (special trains for transporting migrant workers) from May 01 to 06, 2020, to transport more than 100,000 migrant workers within India (PTI, 2020c).

*Implication of Repatriating Migrant Workers on COVID-19 Spread and Transportation Requirements*

Table 8: Dedicated trains and buses required to repatriate migrant workers

| Home states | Migrant workers arriving per day | Maharashtra | | NCT of Delhi | | Gujarat | | Karnataka | | Haryana | | Rest of India | | Total | | Total no. of days |
|---|---|---|---|---|---|---|---|---|---|---|---|---|---|---|---|---|
| | | Trains | Buses | Trains | Buses | Trains | Buses | Trains | Buses | Trains | Buses | Trains | Buses | **Trains** | **Buses** | |
| **UP** | 20,000 | 7 | - | 2 | 60 | 5 | - | 1 | - | 1 | 40 | 7 | 90 | 23 | 190 | 47-71 |
| | 50,000 | 17 | - | 3 | 130 | 11 | - | 2 | - | 3 | 110 | 17 | 230 | 53 | 470 | 19-29 |
| | 100,000 | 34 | - | 6 | 260 | 21 | - | 3 | - | 5 | 210 | 35 | 450 | 104 | 920 | 10-14 |
| **BR** | 20,000 | 4 | - | 3 | - | 4 | - | 1 | - | 3 | - | 12 | 150 | 27 | 150 | 34-51 |
| | 50,000 | 9 | - | 8 | - | 9 | - | 3 | - | 6 | - | 29 | 370 | 64 | 370 | 14-21 |
| | 100,000 | 17 | - | 15 | - | 18 | - | 5 | - | 11 | - | 58 | 740 | 124 | 740 | 7-11 |
| **RJ** | 20,000 | 5 | - | 1 | - | 3 | 120 | 2 | - | 2 | - | 7 | 90 | 20 | 210 | 14-21 |
| | 50,000 | 11 | - | 2 | - | 7 | 300 | 6 | - | 5 | - | 17 | 220 | 48 | 520 | 6-9 |
| | 100,000 | 21 | - | 4 | - | 14 | 600 | 11 | - | 9 | - | 35 | 440 | 94 | 1,040 | 3-5 |
| **MP** | 20,000 | 3 | 110 | 1 | - | 2 | 80 | 1 | - | 1 | - | 9 | 120 | 17 | 310 | 14-21 |
| | 50,000 | 7 | 280 | 2 | - | 5 | 200 | 2 | - | 2 | - | 22 | 290 | 40 | 770 | 6-9 |
| | 100,000 | 13 | 560 | 4 | - | 9 | 400 | 3 | - | 4 | - | 44 | 570 | 77 | 1,530 | 3-4 |
| **MH** | 20,000 | - | - | 1 | - | 4 | 180 | 2 | 80 | 1 | - | 7 | 100 | 15 | 360 | 7-10 |
| | 50,000 | - | - | 2 | - | 10 | 430 | 5 | 190 | 2 | - | 18 | 240 | 37 | 860 | 3-4 |
| | 100,000 | - | - | 3 | - | 20 | 860 | 9 | 380 | 3 | - | 37 | 470 | 72 | 1,710 | 2 |
| **Total** | 20,000 | 19 | 110 | 8 | 60 | 18 | 380 | 7 | 80 | 8 | 40 | 42 | 550 | 102 | 1,220 | - |
| | 50,000 | 44 | 280 | 17 | 130 | 42 | 930 | 18 | 190 | 18 | 110 | 104 | 1,350 | 243 | 2,990 | - |
| | 100,000 | 85 | 560 | 32 | 260 | 82 | 1,860 | 31 | 380 | 32 | 210 | 209 | 2,670 | 471 | 5,940 | - |

*Note:* UP – Uttar Pradesh; BR – Bihar; RJ – Rajasthan; MP – Madhya Pradesh; MH – Maharashtra.

## 6. Conclusions

This study presents the effect of the repatriating migrant workers on the COVID-19 spread in Uttar Pradesh, Bihar, Rajasthan, Madhya Pradesh, and Maharashtra, and the train and bus fleet required to transport the repatriating migrant workers. A modified SEIR model was proposed to estimate the surge in confirmed and active cases of states analyzed. The analysis could reveal the effect of the daily arrival rate of migrant workers on the surge in confirmed and active cases in Uttar Pradesh, Bihar, Madhya Pradesh, and Rajasthan. The surge in confirmed and active cases was expected to be small for lower daily arrival rates, but states may experience disparity in the duration to complete the repatriation. So, a holistic optimization of the repatriation, considering the possible constraints of transportation and health infrastructure, is warranted.

The migrant worker population returning to their home state had a significant impact on the expected number of confirmed and confirmed cases. Since the number of migrant workers returning to Maharashtra was supposedly low, the state's surge in confirmed and active cases was predicted to be smaller than the other states analyzed. As indicated earlier, it could be even lower if the migrant workers leaving a state was considered in the analysis. Comparison of the surge in confirmed and active cases of COVID-19 infection due to the interaction of repatriating migrant workers (see Figure 8) indicated Uttar Pradesh, Bihar, and Madhya Pradesh might need close monitoring. In this study, the migrant workers arriving before May 5, 2020, were neglected. Similarly, migrant workers exhibiting symptoms of COVID-19 after arrival were assumed to be identified on the arrival day. In reality, due to the incubation period of COVID-19, there could be a lag in the symptom exhibition.

As discussed earlier, stringent measures in screening migrant workers before boarding and strict isolation policy after arrival might control the surge in confirmed and active cases. Some of the State Governments planned for 14 to 21 days quarantine period for the arriving migrant workers (Kumar, 2020b). Alternatively, restarting industries with a safe work environment might encourage migrant workers to stay back. The Maharashtra and Karnataka State Governments allowed a few industries, where migrant workers are generally employed, to operate (Poovanna, 2020). Such activities would help the migrant workers to earn a livelihood, reduce the travel demand, and sustain the economy (Sastry, 2020).

This study, instead of considering the effect of variations in the daily travel demand or infection probability with respect to time, adopted different levels of daily travel demand and infection probability. This approach reveals the possible extent of the effect. The surge in COVID-19 infection for any time-dependent variation of input parameters is expected to be within the extent of the effect reported in this study. An SEIR model represents the population in different compartments. Dividing the population into additional compartments to portray the possible interactions between the quarantined individuals and various groups such as health care professionals and essential service providers can yield precise information about the infection spread, but it would require input information that is not readily available. The predicted confirmed and active cases of COVID-19 infected persons greatly depend on both pharmaceutical and non-pharmaceutical infection prevention and control strategies and infection test policies. Any change in these strategies and policies over time would affect the actual surge in confirmed and active cases. Further, the SEIR model parameters such as $\alpha$, $\beta$, $\gamma$, $\delta(t)$, $\lambda(t)$ and $\kappa(t)$ were calibrated for a specific time period, which may change significantly during the period of projections. The data used in the base SEIR model to calibrate the parameters could represent a period of high transmission, which may not sustain long. Hence, there is a possibility that the results presented in the manuscript could be different from the actual surge in confirmed and active COVID-19 cases.

During this study, the non-availability of certain crucial information needed to study the influence of transportation during a biological disaster like COVID-19 was identified. Organized efforts to acquire pertinent information could be helpful in the future. Some of the requirements are summarized as follows:

a) During pandemic like COVID-19, the migrant workers, students, tourists, business executives, family members, etc., who traveled out of town, would attempt to return to their hometown. In India, basic information of migrants settled in a particular state or UT was obtained while collecting the census data. However, this available information was not adequate to estimate the repatriating population. Hence, it is recommended to develop a national-level model to estimate the repatriating population during biological disasters.

b) The number of confirmed and active cases can vary with infection testing rate, test protocol, lockdown policy, etc. This study considered available data till April 29, 2020 for the analysis and did not consider variations of these factors beyond the analysis date. Hence, developing an appropriate model to correlate the testing rate, test protocol, and lockdown policy with COVID-19 spread can help in improving the infection forecasting accuracy.

c) In the absence of realistic data, on arrival, the repatriating migrant workers were assumed to be symptomatic with a higher probability. Interaction of passengers within a confined train compartment and bus coach need to be studied to develop appropriate models for repatriating population of different age group. It is crucial, particularly for the migrant workers and students who are expected to repatriate in large numbers during the biological disaster like COVID-19.

# APPENDIX A

**Table A.1: Average centroid distances between states considered**

| Home states | Distance to migrated states (in km) | | | | | |
|---|---|---|---|---|---|---|
| | Maharashtra | NCT of Delhi | Gujarat | Karnataka | Haryana | Rest of India |
| **Uttar Pradesh** | 1,257 | 554 | 1,399 | 1,730 | 676 | 1,265 |
| **Bihar** | 1,500 | 1,157 | 1,809 | 2,003 | 1,279 | 1,241 |
| **Rajasthan** | 1,040 | 426 | 740 | 1,751 | 359 | 1,573 |
| **Madhya Pradesh** | 618 | 832 | 955 | 1,267 | 916 | 1,165 |
| **Maharashtra** | - | 1,302 | 844 | 583 | 1,361 | 1,199 |

**Table A.2: Additional travel time ($ATT$)**

| Activity | Time (in hours) | |
|---|---|---|
| | Train | Bus |
| Boarding time | 1 | 1 |
| Average delay during travel | 2 | 2 |
| Disinfection time | 2 | 1 |
| Cleaning period | 2 | 1 |
| Change of engine time | 2 | 0 |
| Scheduling delay and random checks | 2 | 6[#] |
| **Total ($ATT$)** | 11 | 11 |

[#]Buses are not scheduled to ply between 12 am, and 6 am.

**Table A.3: Modal split for migrant workers**

| Home states | Mode choice from migrated states, $MCP_{train\ or\ bus}$ (in percentages) | | | | | | | | | | | |
|---|---|---|---|---|---|---|---|---|---|---|---|---|
| | MH | | DL | | GJ | | KR | | HR | | Rest of India | |
| | Train | Bus | Train | Bus | Train | Bus | Train | Bus | Train | Bus | Train | Bus |
| **UP** | 90 | 0 | 45 | 45 | 90 | 0 | 90 | 0 | 45 | 45 | 70 | 20 |
| **BR** | 90 | 0 | 90 | 0 | 90 | 0 | 90 | 0 | 90 | 0 | 70 | 20 |
| **RJ** | 90 | 0 | 90 | 0 | 45 | 45 | 90 | 0 | 90 | 0 | 70 | 20 |
| **MP** | 45 | 45 | 90 | 0 | 45 | 45 | 90 | 0 | 90 | 0 | 70 | 20 |
| **MH** | - | - | 90 | 0 | 45 | 45 | 45 | 45 | 90 | 0 | 70 | 20 |

*Note:* UP – Uttar Pradesh; BR – Bihar; RJ – Rajasthan; MP – Madhya Pradesh; MH – Maharashtra; DL – NCT of Delhi; GJ – Gujarat; KR – Karnataka; HR – Haryana; About 10% migrants are assumed to travel using unorganized travel modes such as hired private vehicles and non-motorized transport modes.

**Steps to estimate train and bus fleet size**

*Assumptions*
Average speed of train = 80 kmph
Average speed of bus = 50 kmph
Average capacity of a train ($C_{train}$) = 1200 passengers
Average capacity of a bus ($C_{bus}$) = 25 passengers (at 50% of the standard capacity)
Average distance between states is available in Table A1.

Additional travel time required during COVID-19 is available in Table A2.
Percentage of migrant workers choosing a specific travel mode ($MCP_{train\ or\ bus}$) is available in Table A3.
Percentage of migrant workers repatriating from a migrated state to a home state (i.e., $MWP^{MS\ to\ HS}$) is available in Table 6.
Trains and buses were assumed to return empty in the roundtrip.
Buses would halt for 0.5 hours (i.e., *halt duration*) after every 4 hours of drive (i.e., *the time between halts*) to facilitate bio-break.
Buses travel till the common state borders.

*Step 1:* Estimate roundtrip travel time ($RTT$) in hours for train and bus using equation A.1.
$$RTT_{train\ or\ bus} = \frac{2 \times Distance\ between\ states}{Average\ speed\ or\ train\ or\ bus} \tag{A.1}$$

*Step 2:* Estimate overall travel time ($OTT$) in days for train and bus using equations A.2 and A.3, respectively.
$$OTT_{train} = \frac{(RTT_{train} + ATT_{train})}{24} \tag{A.2}$$

$$OTT_{bus} = \frac{\left(\frac{RTT_{bus}}{2} + ATT_{bus}\right) + \left(\frac{RTT_{bus}}{2 \times time\ between\ halts}\right) \times halt\ duration}{24} \tag{A.3}$$

*Step 3:* Estimate the number of migrant workers repatriating per day from a migrated state to their home state ($NMW^{MS\ to\ HS}$) using equation A.4.
$$NMW^{MS\ to\ HS} = Daily\ arrival\ rate \times MWP^{MS\ to\ HS} \tag{A.4}$$

*Step 4:* Estimate the train and bus fleet size to repatriate migrant workers from migrated state to home state ($FS^{MS\ to\ HS}$) using equation A.5.
$$FS^{MS\ to\ HS}_{train\ or\ bus} = \left\{\frac{NMW^{MS\ to\ HS} \times MCP_{train\ or\ bus}}{C_{train\ or\ bus}}\right\} \times OTT_{train\ or\ bus} \tag{A.5}$$